\documentclass[openany]{article}

\usepackage[english]{babel}
\usepackage{commands}
\usepackage{xcolor}
\usepackage{authblk}
\usepackage{soul}

\usepackage{fancyhdr}	

\title{A compartmental model for \textit{Xylella fastidiosa} diseases with explicit vector seasonal dynamics}

\author[1]{Alex Giménez-Romero}
\author[2]{Eduardo Moralejo}
\author[1]{Manuel A. Matías}

\affil[1]{Instituto de Física Interdisciplinar y Sistemas Complejos (IFISC, CSIC-UIB), Campus UIB, 07122 Palma de Mallorca, Spain.}
\affil[2]{Tragsa, Passatge Cala Figuera 6, 07009 Palma de Mallorca, Spain.}

\date{}

\begin{document}

\maketitle

\begin{abstract}

 The bacterium \textit{Xylella fastidiosa} (Xf) is mainly transmitted by the spittlebug, \textit{ Philaenus spumarius}, in Europe, where it has caused significant economic damage to olive and almond trees. Understanding the factors that determine disease dynamics in pathosystems that share similarities can help design control strategies focused on minimizing transmission chains. Here we introduce a compartmental model for Xf-caused diseases in Europe that accounts for the main relevant epidemiological processes, including the seasonal dynamics of \textit{P. spumarius}. The model was confronted with epidemiological data from the two major outbreaks of Xf in Europe, the olive quick disease syndrome (OQDS) in Apulia, Italy, caused by the subspecies \textit{pauca}, and the almond leaf scorch disease (ALSD) in Majorca, Spain, caused by subspecies \textit{multiplex} and \textit{fastidiosa}. Using a Bayesian inference framework, we show how the model successfully reproduces the general field data in both diseases. In a global sensitivity analysis, the vector-plant and plant-vector transmission rates, together with the vector removal rate, were the most influential parameters in determining the time of the infected host population peak, the incidence peak and the final number of dead hosts. We also used our model to check different vector-based control strategies, showing that a joint strategy focused on increasing the rate of vector removal while lowering the number of annual newborn vectors is optimal for disease control.
 
\end{abstract}

\vspace{0.5cm}

\textbf{Keywords:} Epidemiology, Population Biology, Modeling.

\vspace{1cm}

    Mathematical and computational modeling in Ecology and, in particular, Epidemiology have been recently recognized as powerful approaches to guide empirical work and provide a framework for the synthesis, analysis and development of conservation plans and policy-making (\cite{levin1992mathematics,murray1989mathematical,sarkar2006biodiversity,Chew2014}). Plant epidemics, mainly plant-virus diseases, have been often described by compartmental models, which deal with the overriding importance of transmission mechanisms in determining epidemic dynamics (\cite{Jeger1998,Jeger2004,Madden2000}). These models have contributed to providing answers to some questions related to the ecology of plant diseases and have led to direct applications in disease control while guiding research directions (\cite{Jeger2019}).

    The emergence of vector-borne plant pathogens in new areas causing huge economic impacts, such as \textit{Xylella fastidiosa} and the \textit{Candidatus} Liberibacter spp. (Huanglongbing or citrus greening), has sparked interest in modeling vector-transmitted plant-disease epidemics (\cite{chiyaka2012modeling,Jeger2019}). The vector-borne bacterium \textit{X. fastidiosa} (Xf) is a multi-host pathogen endemic to the Americas that causes economically important diseases, mostly in woody crops (\cite{Hopkins2002}). Xf is a genetically diverse species with three evolutionary well-defined clades forming the \textit{pauca}, \textit{fastidiosa}, and \textit{multiplex} subspecies, native from South, Central, and North America, respectively (\cite{vanhove2019genomic}). Within each subspecies, diverse genetic lineages with different host ranges are found. Xf is transmitted non-specifically by xylem-sap-feeding insects belonging to the sharpshooter leafhoppers (Hemiptera: Cicadellidae, Cicadellinae) and spittlebugs (Hemiptera: Cercopoidae) (\cite{Redak2004}).

    Recently, Xf has gained renewed interest due to the massive mortality of olive trees in Apulia, Italy (\cite{saponari2019xylella}). The first focus of the olive quick decline syndrome (OQDS) was detected in 2013 around Gallipoli (\cite{saponari2013identification}) and since then has spread throughout the region by the meadow spittlebug, \textit{Philaenus spumarius}. Although this was the first official detection of Xf in Europe, it has recently been demonstrated that the pathogen arrived much earlier in Corsica (\cite{Soubeyrand2018}) and the Balearic islands (\cite{Moralejo2020}). Around 1993, two strains of the subspecies \textit{fastidiosa} (ST1) and \textit{multiplex} (ST81) were introduced from California to Mallorca (Spain) with infected almond plants (\cite{Moralejo2020}). To date, over 80\% of the almond trees in Mallorca show leaf scorch symptoms and the outbreak has changed the iconic rural landscape of this Mediterranean island (\cite{Olmo2021b}).  
    
    Several epidemic models have been already developed for Xf-diseases, but they lack a realistic description of some relevant processes (\cite{Jeger2019}). Furthermore, some of these models assume a simple general form for infected host dynamics (\cite{White2017,Abboud2019,Daugherty2019}) or use a simplified S-I compartmental scheme for hosts, disregarding important features such as the latent period or the host mortality rate (\cite{Soubeyrand2018}). Models that do take these features into account, however, do not explicitly model the population of vectors responsible for disease transmission (\cite{White2020}). Other more recent models have taken a step further in explicitly modeling the vector population (\cite{BRUNETTI2020}), but the characterization of its dynamics is still relatively simple, as it overlooks the known seasonal patterns of vector abundance. Thus, there is a need to continue advancing in the modeling of Xf diseases by developing more realistic models that can elucidate the fundamental processes involved in vector-host-pathogen interactions and help to design effective control strategies. 
    
    In this work, we develop a deterministic continuous-time compartmental model to describe the general epidemiological dynamics of diseases produced by Xf in Europe, explicitly accounting for key biological aspects of the main vector \textit{P. spumarius}, including its seasonal dynamics. Our model is able to describe field data from the two major European outbreaks: the olive quick disease syndrome (OQDS) in Apulia, Italy, caused by the subspecies \textit{pauca}, and the almond leaf scorch disease (ALSD) in Majorca, Spain, caused by subspecies \textit{multiplex} and \textit{fastidiosa}. We aimed to find the most influential parameters in the model with respect to incidence and mortality in both diseases by performing a global sensibility analysis. With this information, the next goal was to explore control strategies acting especially on the vector population.

\section*{MATERIALS AND METHODS}
    
\subsection*{Epidemic model: the SEIR-V model}

    We developed a deterministic continuous-time compartmental model that incorporates the specific biological features of Xf diseases in Europe, including the dynamics of the main relevant vector \textit{P. spumarius} (\cite{ Cavalieri2019}). To build the model we took the following considerations: (i) we assume there is no winter recovery of infected hosts and thus they die sometime after infection; (ii) hosts show an asymptomatic period in which they are non-infectious in practice (exposed compartment) because the bacteria are not yet systemically extended (\cite{teviotdale2003almond,Stevenson238}), while vectors are infectious immediately after acquiring the bacterium (\cite{Fierro2019}); (iii) vectors have an annual life cycle without mother-to-offspring disease transmission (\cite{freitag1951host,purcell1979evidence}), so we consider the annual emergence of susceptible newborn vectors and a constant death rate for both susceptible and infected vectors; (iv) infected vectors carry the bacterium during their entire lifespan without affecting their fitness ; and finally, (v) we do not consider host recruitment or natural death given that the typical development time of Xf-epidemics is faster than the typical host's life cycle.
    
\begin{figure}[H]
        \centering
        \includegraphics[width=0.8\textwidth]{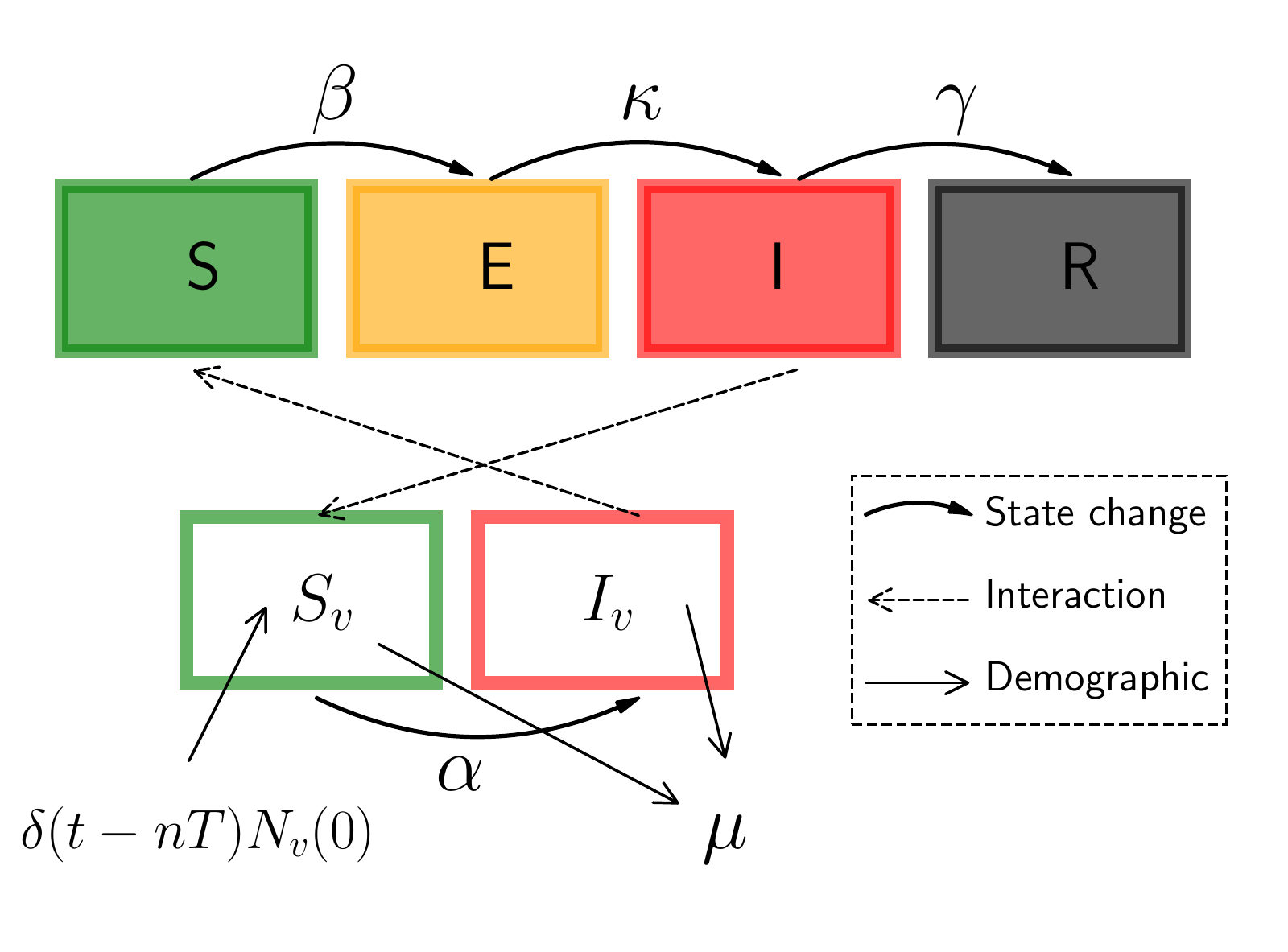}
        \caption{Schematic representation of the model \cref{eq:SEIR_v}. Boxes are the compartments in which the population is divided, solid curved arrows represent changes in state, i.e. transitions between compartments, dashed arrows depict the crossed interaction between hosts and vectors and solid straight arrows represent demographic changes in vector population.}
        \label{fig:model_diagram}
    \end{figure}

    Altogether, our deterministic continuous-time compartmental model consists of six compartments, four describing the host population (susceptible, $S_H$, exposed, $E_H$, infected, $I_H$, and removed, $R_H$), and two describing the vector population (susceptible, $S_V$, and infected, $I_V$). The model is defined according to the following processes,
    \begin{equation}\label{eq:scheme_infection}
        \begin{aligned}
            &S_H+I_V \stackrel{\beta}{\rightarrow} E_H + I_V, \quad
            E_H \stackrel{\kappa}{\rightarrow} I_H, \quad
            I_H  \stackrel{\gamma}{\rightarrow} R_H \\
            &S_V+I_H \stackrel{\alpha}{\rightarrow} I_V+I_H, \quad S_V \stackrel{\mu}{\rightarrow} \varnothing, \quad I_V \stackrel{\mu}{\rightarrow} \varnothing
            \ ,
        \end{aligned}   
    \end{equation}
    which are illustrated in \cref{fig:model_diagram}, being the birth of new susceptible vectors described as a source term.
    Thus, the host-vector compartmental model is written as,
    \begin{equation}\label{eq:SEIR_v}
        \begin{aligned}
            \dot{S}_H &=-\beta S_H I_v / N_H \\
            \dot{E}_H &=\beta S_H I_v / N_H - \gamma I_H \\
            \dot{I}_H &=\kappa E - \gamma I \\
            \dot{R}_H &=\gamma I_H \\
            \dot{S}_v &= N_v(0)\sum_{n=1}^{\infty}\delta(t-nT) -\alpha S_v I_H / N_H - \mu S_v \\
            \dot{I}_v &=\alpha S_v I_H / N_H - \mu I_v \ .
        \end{aligned}
    \end{equation}

    The model describes the exposure of susceptible hosts, $S_H$, at a rate $\beta$ through their interaction with infected vectors, $I_v$, while susceptible vectors, $S_v$, get infected immediately at a rate $\alpha$ through their interaction with infected hosts $I_H$. Exposed hosts get infected at rate $\kappa$, being the mean latent period $\tau_E=1/\kappa$, while infected hosts die at rate $\gamma$, having a mean infectious period of $\tau_I=1/\gamma$. Infected vectors stay infected for the rest of their lifetime. Regarding the seasonal dynamics of vectors, we assume that new adults emerge synchronously each year in fields being all susceptible. This is represented by the term $N_v(0)\sum_{n=1}^{\infty}\delta(t-nT)$ in \cref{eq:SEIR_v}, where $T=\SI{1}{yr}$ is the period and $\delta(t-nT)$ is Dirac delta function, with the property: $\delta(x-x_0)=0$ for all $x\neq x_0$ and $\int g(t)\delta(t-t_0)\dif t=g(t_0)$. Vectors die or exit the field at a given rate $\mu$, which we consider identical for susceptible and infected vectors. For simplicity, we consider that the quantity of annual newborn adults, $N_v(0)$, is constant. This outburst of new adults followed by an exponential decay resembles the temporal patterns on the abundance of \textit{P. spumarius} observed in crop fields  (\cite{Antonatos2021,Beal2021,Cornara2017,Lopez2021}) (see \cref{fig:vector_dynamics}).
  
    In \cref{eq:SEIR_v} the crossed nonlinear terms, $S_H I_v$ and $S_v I_H$ are written divided by the total host population, $N_H$. Thus, the plant-to-vector infection process is modeled using standard incidence, which is frequency dependent, while the vector-to-plant infection process is modeled using mass action incidence, which is density dependent (\cite{MartchevaBook}). This implies that doubling the number of vectors in the crop field would double the number of resulting infected hosts, as this process is population-dependent, while doubling the number of hosts would not result in more vectors being infected, as this process only depends on the contact probability. We think this is the most reasonable assumption because increasing the number of hosts is expected to also increase the area of the field, while the number of vectors is independent.
    
\subsection*{Basic reproductive number}

    The basic reproductive number, $R_0$, of the model cannot be trivially computed using standard methods such as the Next Generation Matrix (NGM) (\cite{Diekmann2010}), as there is no pre-pandemic fixed point in the system of differential equations \cref{eq:SEIR_v}. For periodic vector populations, rigorous methods have been developed (\cite{Bacaer2007}) but not for the case of growing or decaying vector populations. Here we use the simple method developed in the work of \cite{Gimenez2022} (see \cref{app:R0}), which effectively computes the average number of secondary infections produced by an initially infected individual in one generation. Thus, the effective basic reproductive number is given by
    \begin{equation}\label{eq:R0}
        R_0=\frac{\beta\alpha}{\mu\gamma}\frac{S_H(0)}{{N_H}^2}\frac{N_v(0)}{\mu\tau}\left(1-e^{-\mu\tau}\right) \ ,
    \end{equation}
    where $\tau$ corresponds to the time length of one generation, in our case one year. This $R_0$ is calculated using the initial susceptible host population, $S_H(0)$. Below we will also use a time-dependent $R_0(t)$ using $S_H(t)$.
    
\subsection*{Epidemiological data}

    Epidemiological data from an ALSD outbreak in the island of Mallorca Balearic Islands, Spain were taken from \cite{Moralejo2020}. Dated phylogenetic analysis and estimates of disease incidence showed that the introduction of both subspecies occurred around 1993 and $\sim 79$\% almond trees were infected in 2017 (\cite{Moralejo2020}). The annual proportion of infected individuals in the almond tree population between 1993 and 2017 was estimated by analyzing through qPCR the presence of Xf-DNA in the growth rings of $34$ sampled trees (cf. Fig. 3 in (\cite{Moralejo2020})). The disease progression curve was estimated without distinguishing whether infections were caused by \textit{multiplex} or \textit{fastidiosa} subspecies. In addition, a two-sided bootstrap confidence interval for each data point was set using the SciPy bootstrap function in Python (\cite{SciPy}). On the other hand, epidemic data for OQDS were retrieved from (\cite{White2020}). The data consisted of 2 to 3 yearly censuses of symptom prevalence in 17 olive groves infected with Xf subsp. \textit{pauca} in Apulia, Italy, which were aggregated to fit our model as shown in Fig. 4 in (\cite{White2020}). Because the compartments of our model are not in one-to-one correspondence with those shown in the work of White et al. (\cite{White2020}), we used the sum of the symptomatic and desiccated infected trees in the dataset ($I_S+I_D$) to fit the sum of the infected and dead trees ($I+R$) and the sum of susceptible and asymptomatic hosts ($S+I_A$) to fit the sum of susceptible and exposed hosts ($S+E$). The processed data used to fit the model can be found in (\cite{CODE}), while the raw data can be found in the supplementary data accessible online of the cited articles (\cite{Moralejo2020,White2020}).
    
\subsection*{Model fitting through Bayesian Inference}
    
    We employed an informative normal $\mathcal{N}(\hat{\mu},\hat{\sigma}^2)$ prior distribution, with $\hat{\mu}$ and 
    $\sigma$, the mean and standard deviation, respectively, for previously measured parameters in the literature, such as the infected and latent periods for ALSD, $\tau_I\sim\mathcal{N}(14, 4)$, $\tau_E\sim\mathcal{N}(4, 1)$ (\cite{teviotdale2003almond, Moralejo2020}) and OQDS, $\tau_I\sim\mathcal{N}(3.5, 1)$, $\tau_E\sim\mathcal{N}(1.75, 0.5)$ (\cite{Fierro2019}). The corresponding rates are given by $\gamma=1/\tau_I$ and $\kappa=1/\tau_E$, respectively. Similarly, a prior normal distribution was used for the removal rate of vectors, $\mu\sim\mathcal{N}(0.02, 0.0075)$, as the mean value $\mu=0.02$ already captures the vector dynamics observed in field-data (\cref{fig:vector_dynamics}). Regarding the prior distribution for the transmission rates a very wide and uninformative uniform prior distribution, $\beta\sim \mathcal{U}(0.001, 1)$ and $\alpha\sim\mathcal{U}(0.001, 1)$, was used for each parameter. The number of hosts, $N_H$, was already provided in the datasets, while, given the lack of information about the vector population, we assumed $N_v(0)=N_H/2$ for the initial vector population of each year. However, we tested the robustness of our results by changing $N_v(0)$.
    
    The posterior distributions of the parameters were approximated using the Markov Chain Monte Carlo algorithm No U-Turn Sampler (NUTS) with the recommended target acceptance rate of 65\% (\cite{Homan2014}). To ensure a proper convergence, we constructed three independent Markov Chains with $10^5$ iterations each after a burn-in of $10^4$ iterations and checked that the results were statistically equivalent. For each chain, we started at the maximum-likelihood parameters yielded by the Nelder-Mead algorithm with 1000 iterations.

    The parameters of our compartmental model were determined by fitting the model to data by means of a Bayesian Inference framework using the Turing.jl package (\cite{Turing.jl}) in Julia (\cite{julia}). The scripts used to fit the model can be found in (\cite{CODE}).
    
\subsection*{Sensitivity Analysis}
    
    We performed a Global Sensitivity Analysis (GSA) (\cite{Sensitivity_analysis_book}) of the model to assess the relative contribution of its parameters and their interactions with different features of the epidemic. In contrast to the Local Sensitivity Analysis (LSA), the GSA assesses the influence of a large domain of the parameter space in the desired outputs of the model. We performed GSA by means of a variance-based analysis, the Sobol method (\cite{SOBOL2001271}). This particular method provides information not only on how a particular parameter alone influences the model outputs (as happens with LSA), but also due to the nonlinear interactions among two or more parameters. Briefly, the method considers the model output, $Y$, as a general function of the inputs, $f(x_1, ..., x_n)$, so that the variance of the output, $Var(Y)$ is decomposed as the sum of the variances given by the variations of the parameters alone and its interactions, $Var(Y)=\sum_{i=1}^nVar(f(x_i)) + \sum_{i<j}^nVar(f(x_i, x_j)) + \cdots$. This information is organized in what are known as Sobol indices. The total order indices are a measure of the total variance of the output quantity caused by variations of the input parameter and its interactions, $S_T=Var(f(x_1,...,x_n))/Var(Y)$. First order (or ``main effect'') indices are a measure of the contribution to the output variance given by the variation of the parameter alone, but averaged over the variations in other input parameters, $S_i=Var(f(x_i))/Var(Y)$. Second-order indices take into account first-order interactions between parameters, $S_{ij}=Var(f(x_i,x_j)) / Var(Y)$. Further indices can be obtained, describing the influence of higher-order interactions between parameters, but these are not going to be considered.
    
    Following the Sobol method, we analyzed the variation of the time at which the infected population peaks, $t_{peak}$, the magnitude of this peak, $I_{peak}$ and the final number of dead hosts, $R_{\infty}$, relative to variations of the model parameters. The method was implemented within the Julia high-level programming language (\cite{julia}) using the sub-package DiffEqSensitivity.jl in DifferentialEquations.jl package (\cite{DifferentialEquations.jl}).

\section*{RESULTS}

\subsection*{Model fit and parameter estimates}

\begin{figure}[H]
        \centering
        \includegraphics[width=\textwidth]{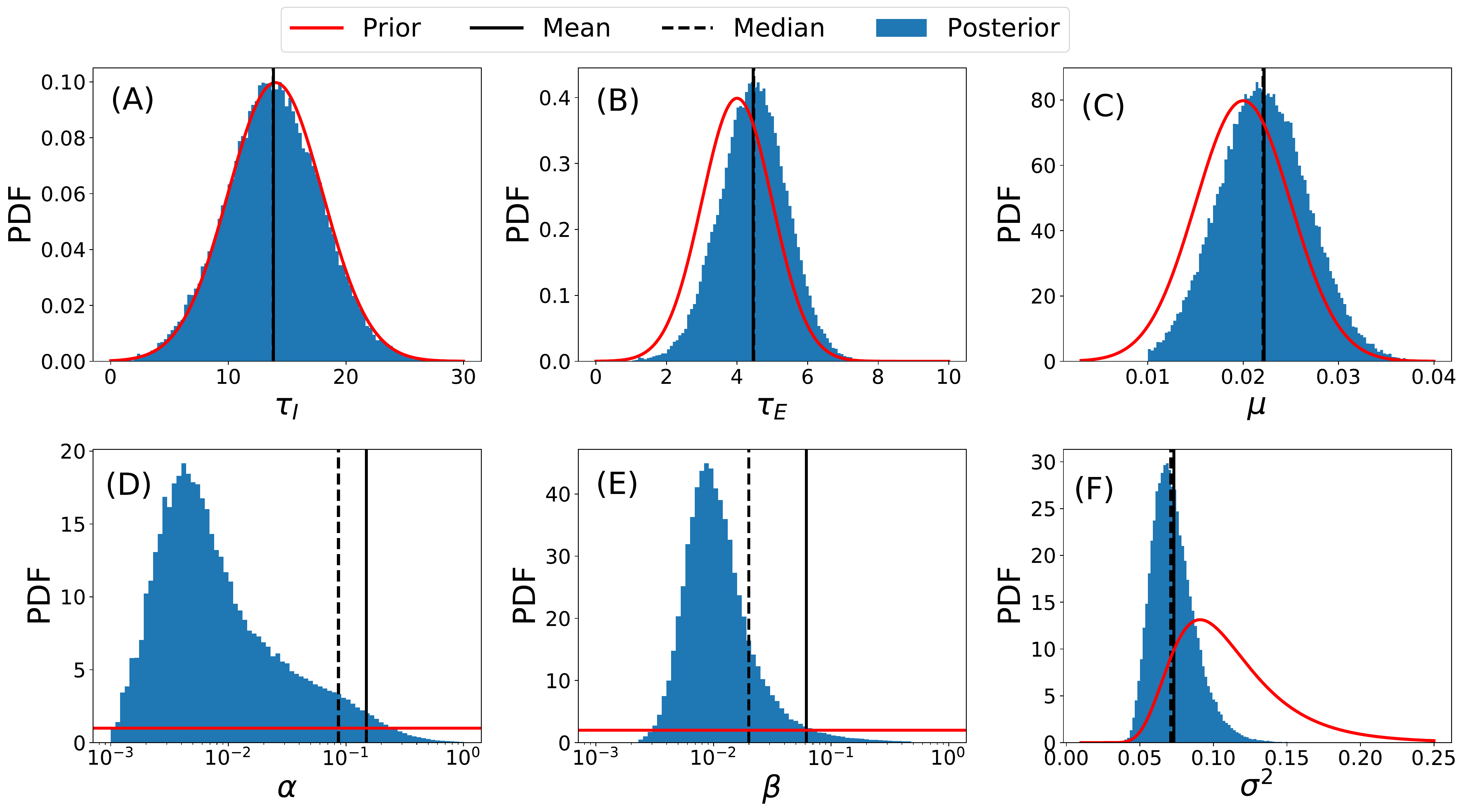}
        \caption{Posterior (blue histograms) and prior (red line) distributions of the model parameters for ALSD. Solid and dashed black lines correspond to the mean and median of the posterior distributions. (A) Host infected period $\tau_I=1/\gamma$. (B) Host latent period $\tau_E=1/\kappa$.  (C) Vector removal rate $\mu$. (D) Vector infection rate $\alpha$. (E) Host infection rate $\beta$. (F) The variance of the field data $\sigma^2$.}
        \label{fig:parameter_estimates_ALSD}
    \end{figure}

\begin{figure}[H]
        \centering
        \includegraphics[width=\textwidth]{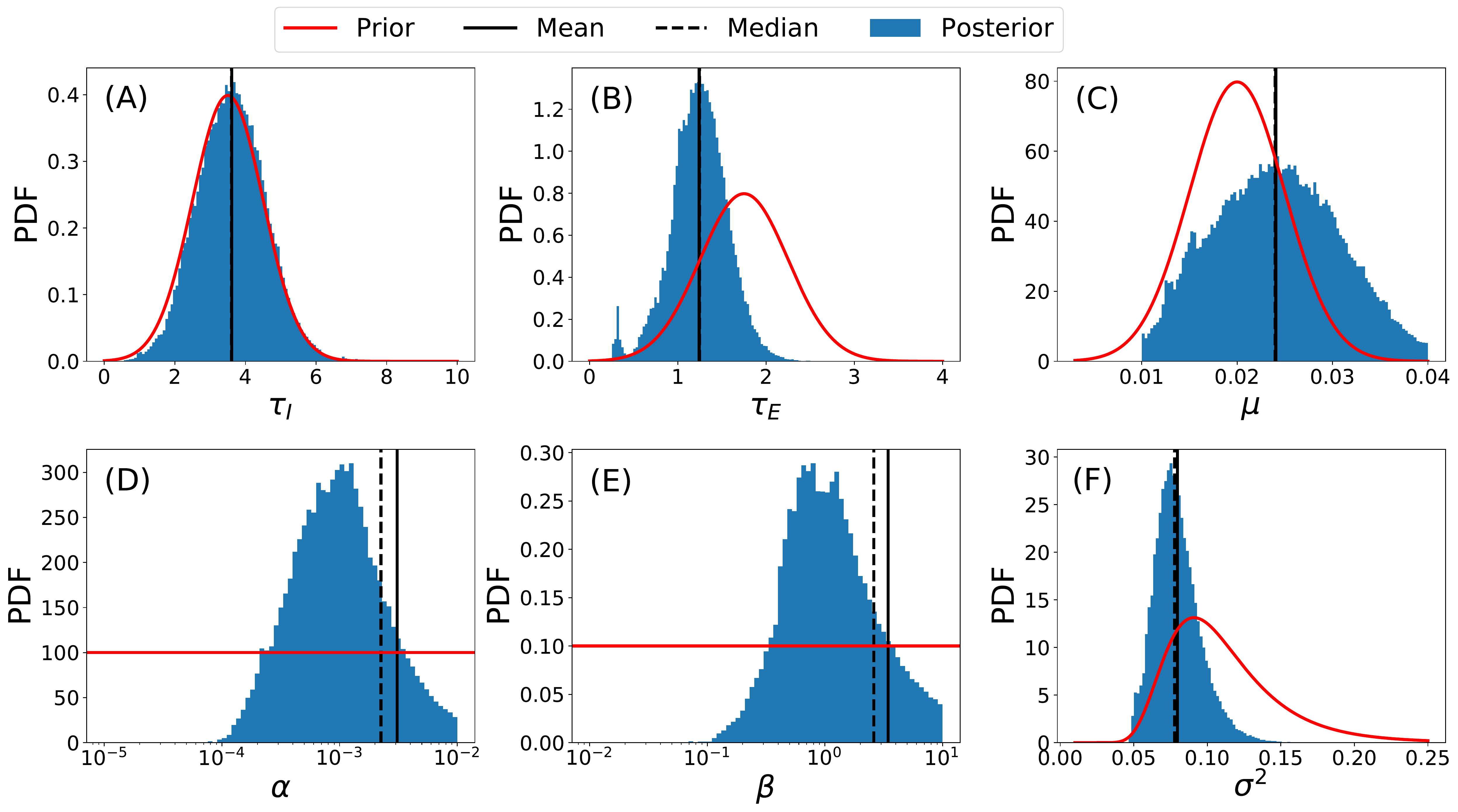}
        \caption{Posterior (blue histograms) and prior (red line) distributions of the model parameters for OQDS. Solid and dashed black lines correspond to the mean and median of the posterior distributions. (A) Host infected period $\tau_I=1/\gamma$. (B) Host latent period $\tau_E=1/\kappa$. (C) Vector removal rate $\mu$. (D) Vector infection rate $\alpha$. (E) Host infection rate $\beta$. (F) Variance of the field data $\sigma^2$.}
        \label{fig:parameter_estimates_OQDS}
    \end{figure}

    The posterior distributions of the fitted parameters including their estimated mean and median for ALSD and OQDS are shown in \cref{fig:parameter_estimates_ALSD,fig:parameter_estimates_OQDS}, respectively, together with the assumed prior distributions. We observe that the literature-driven priors for the latent and infected period, $\tau_E$ and $\tau_I$, were already very good guesses and changed slightly converging to the appropriate distribution that better fitted the epidemic data for both ALSD and OQDS (\cref{fig:parameter_estimates_ALSD}(A-B) and \cref{fig:parameter_estimates_OQDS}(A-B)). Similarly, the prior for the vector removal rate, $\mu$, obtained from field data, was good enough so that little changes were needed for convergence (\cref{fig:parameter_estimates_ALSD}(C) and \cref{fig:parameter_estimates_OQDS}(C)). On the other hand, we also observe that the completely uninformative priors for the transmission rates successfully converged to the posterior distributions (\cref{fig:parameter_estimates_ALSD}(D-E) and \cref{fig:parameter_estimates_OQDS}(D-E)). 

\begin{figure}[H]
        \centering
        \includegraphics[width=0.9\textwidth]{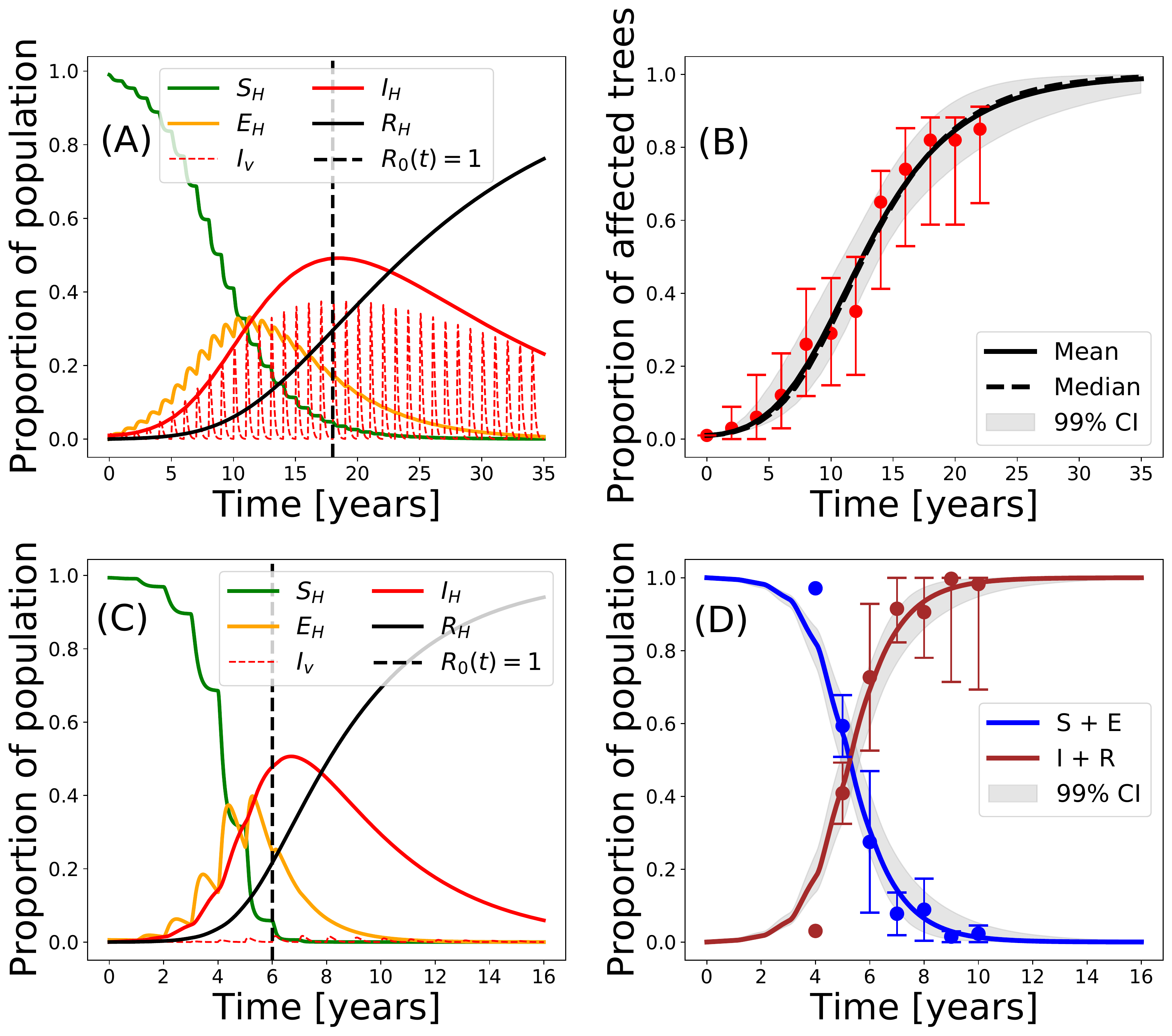}
        \caption{(A) Simulation of the model with the best-fit parameters for ALSD. (B) Model fit to field data by means of the mean and median values of the posterior distributions of the parameters for ALSD. (C) Simulation of the model with the best-fit parameters for OQDS. (D) Model fit to field data by means of the mean and median values of the posterior distributions of the parameters for OQDS. The gray-shaded area corresponds to the 99\% confidence interval. The error bars for the field data correspond to their 95\% confidence interval obtained with a bootstrapping technique.}
        \label{fig:best_fit_model}
    \end{figure}

    The latter distributions are far from a Gaussian-like shape (note that the x-axis is log-scaled), being heavy-tailed. This kind of distribution highly distorts the statistical measures of mean, median and standard error, indicating that the estimates for transmission rates are not as robust as the estimates for the other parameters. These rather uninformative distributions are probably arising because of the lack of data about the vector, i.e. $S_v(t)$ and $I_v(t)$, to constrain the fits. In essence, many combinations of $\alpha$ and $\beta$ can similarly fit the host data while yielding quite different time series for $S_v(t)$ and $I_v(t)$, which cannot be contrasted due to the lack of field data. Nevertheless, the obtained best-fit mean and median parameters, although quite different, are able to perfectly fit the data (\cref{fig:best_fit_model}). Finally, we also observe that the variance for the field data also converged to a bell-shaped distribution.
    
    Mean and median parameter estimates, i.e. the best-fit parameter values for ALSD and OQDS, are summarized in \cref{tab:parameter_estimates_ALSD,tab:parameter_estimates_OQDS}, respectively. As already seen from the posterior distributions, the best-fit values for $\tau_E$, $\tau_I$ and $\mu$ are close to the ones given by literature and field data for both diseases. Conversely, $\alpha$ and $\beta$ are rather uninformative, as their 95\% confidence intervals cover almost two orders of magnitude. This again indicates that without some data about the evolution of the vector states in time, $S_v(t)$ and $I_v(t)$, it is nearly impossible to find the proper values for these parameters.
    
    \begin{table}[H]
        \centering
        \caption{Estimated epidemiological parameters from Bayesian model fitting to the disease progression curve of ALSD in Mallorca.}
        \resizebox{\textwidth}{!}{
        \begin{tabular}{cccccc}
            \hline \hline
            \textbf{Parameter} & \textbf{Definition} & \textbf{Units} & \textbf{Posterior Mean} & \textbf{Posterior Median} & \textbf{95\% C.I.} \\ \hline
            $\tau_I$ & Host infected period & yr & $13.84$ & $13.82$ & $[7.12, 20.47]$ \\
            $\tau_E$ & Host latent period & yr & $4.46$ & $4.47$ & $[2.88, 5.99]$ \\
            $\beta$  & Host infection rate & $\SI{}{\frac{\#host}{\#vector \cdot day}}$ & $0.062$ & $0.02$ & $[0.0061, 0.3013]$ \\
            $\alpha$ & Vector infection rate & $\SI{}{day^{-1}}$ & $0.15$ & $0.086$ & $[0.0047, 0.54]$ \\
            $\mu$ & Vector removal rate & $\SI{}{day^{-1}}$ & $0.0222$ & $0.0221$ & $[0.015, 0.030]$ \\
            $R_0$ & Basic reproductive number & - & 133 & 25 & - \\ \hline \hline
        \end{tabular}
        }
        \label{tab:parameter_estimates_ALSD}
    \end{table}
    
    \begin{table}[H]
        \centering
        \caption{Estimated epidemiological parameters from Bayesian model fitting to the disease progression curve of OQDS in Apulia.}
        \resizebox{\textwidth}{!}{%
        \begin{tabular}{cccccc}
            \hline \hline
            \textbf{Parameter} & \textbf{Definition} & \textbf{Units} & \textbf{Posterior Mean} & \textbf{Posterior Median} & \textbf{95\% C.I.}\\ \hline
            $\tau_I$ & Host infected period & yr & $3.61$ & $3.60$ & $[2.06, 5.20]$ \\
            $\tau_E$ & Host latent period & yr & $1.24$ & $1.25$ & $[0.70, 1.75]$ \\
            $\beta$  & Host infection rate & $\SI{}{\frac{\#host}{\#vector \cdot day}}$ & $3.44$ & $2.60$ & $[0.55, 8.79]$\\
            $\alpha$ & Vector infection rate & $\SI{}{day^{-1}}$ & $0.0031$ & $0.0022$ & $[0.0005, 0.0084]$ \\
            $\mu$  & Vector removal rate & $\SI{}{day^{-1}}$ & $0.0240$ & $0.0240$ & $[0.014, 0.035]$ \\ 
            $R_0$ & Basic reproductive number & - & 33 & 21 & - \\ \hline \hline
        \end{tabular}
        }
        \label{tab:parameter_estimates_OQDS}
    \end{table}
    
    Overall, the estimated mean and median parameters consistently fit the provided data within the 99\% confidence interval for the ALSD and OQDS outbreaks (\cref{fig:best_fit_model}(B,D)). We also computed the instantaneous reproductive number, $R_0(t)$, by using \cref{eq:R0} with $S_H(t)$ instead of only $S_H(0)$ along the simulation. Noteworthy, $R_0(t)=1$ coincides with the stopping of new infections being produced, i.e. the number of exposed hosts does not increase (\cref{fig:best_fit_model}(A,C)). This supports our approximate method for computing the reproductive number for Xf diseases (\cref{app:R0}, \cref{eq:R0}). Due to the different time scales of both epidemics ($\tau_I^{ALSD}+\tau_E^{ALSD} > \tau_I^{OQDS}+\tau_E^{OQDS}$), the OQDS outbreak dies out earlier than the one for ALSD.
    
    We notice that for ALSD a large proportion of the vector population gets infected every year (\cref{fig:best_fit_model}(A)), while a very small proportion is needed in OQDS to produce a lethal outbreak (\cref{fig:best_fit_model}(C)). However, this last statement is rather unrealistic, as around 50\% of the vectors that are captured in Apulia are indeed infected by Xf (\cite{Cavalieri2019,cornara2017transmission}). Thus, the evolution of the infected vector population should be qualitatively similar to that obtained for ALSD (\cref{fig:best_fit_model}(C)). As previously explained, different suitable values of these parameters should give rise to similar progression curves for the hosts while different ones for the vectors cannot be contrasted due to the lack of data. 
    
\begin{figure}[H]
        \centering
        \includegraphics[width=\textwidth]{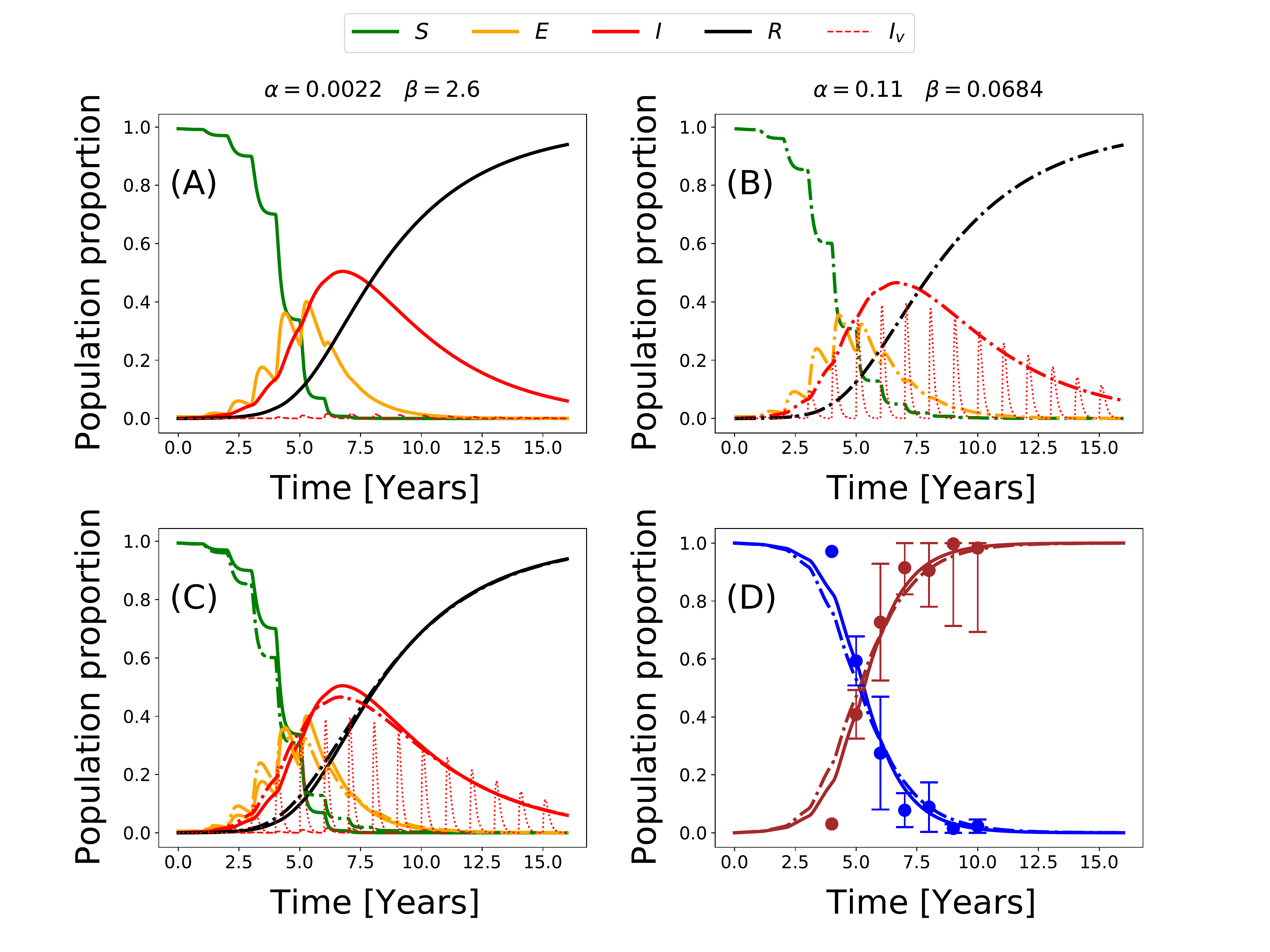}
        \caption{(A) Simulation of the model with the original best-fit parameters for OQDS. (B) Simulation of the model with the original best-fit parameters for OQDS but with different $\alpha$, $\beta$ values. (C) Comparison of the progression curves. Note that the curves for the hosts are very similar while the curve for the infected vector population is very different. (D) Comparison of the model fit to the data with both simulations. Solid lines correspond to results with the original best-fit parameters while dash-dot lines correspond to the results of the more realistic scenario with different $\alpha$ and $\beta$.}
        \label{fig:best_fit_model_OQDS}
    \end{figure}

    Nevertheless, by manually exploring other values for $\alpha$ and $\beta$ parameters, we can obtain a more biologically plausible scenario for the OQDS that is still able to fit the available data for the hosts. \cref{fig:best_fit_model_OQDS}(A) shows a simulation of the model with previously inferred best-fit median parameters for OQDS. By changing the values of $\alpha$ and $\beta$, we obtain a more realistic scenario, i.e. around a 50\% of the vector population getting infected during the outbreak (\cref{fig:best_fit_model_OQDS}(B)) (\cite{Cavalieri2019, cornara2017transmission}). This change in the transmission parameters only affects the progression curve of the infected vector population, being the progression of the host compartments practically unchanged (\cref{fig:best_fit_model_OQDS}(C)). Anyway, both sets of parameter values for $\alpha$ and $\beta$can properly fit the field data (\cref{fig:best_fit_model_OQDS}(D)).

    The model adjusted to the progression curves of both diseases indicates that the transmission rate $\alpha$ must be greater than $\beta$ when the proportion of infected vectors is relatively high ($>30\%$). We checked if the relation between $\alpha$ and $\beta$ held when changing the assumed $N_v(0)=N_H/2$, obtaining that it kept approximately the same for very different values of the initial vector population.

\subsection*{Global Sensitivity Analysis}
    
\begin{figure}[H]
        \centering
        \includegraphics[width=\textwidth]{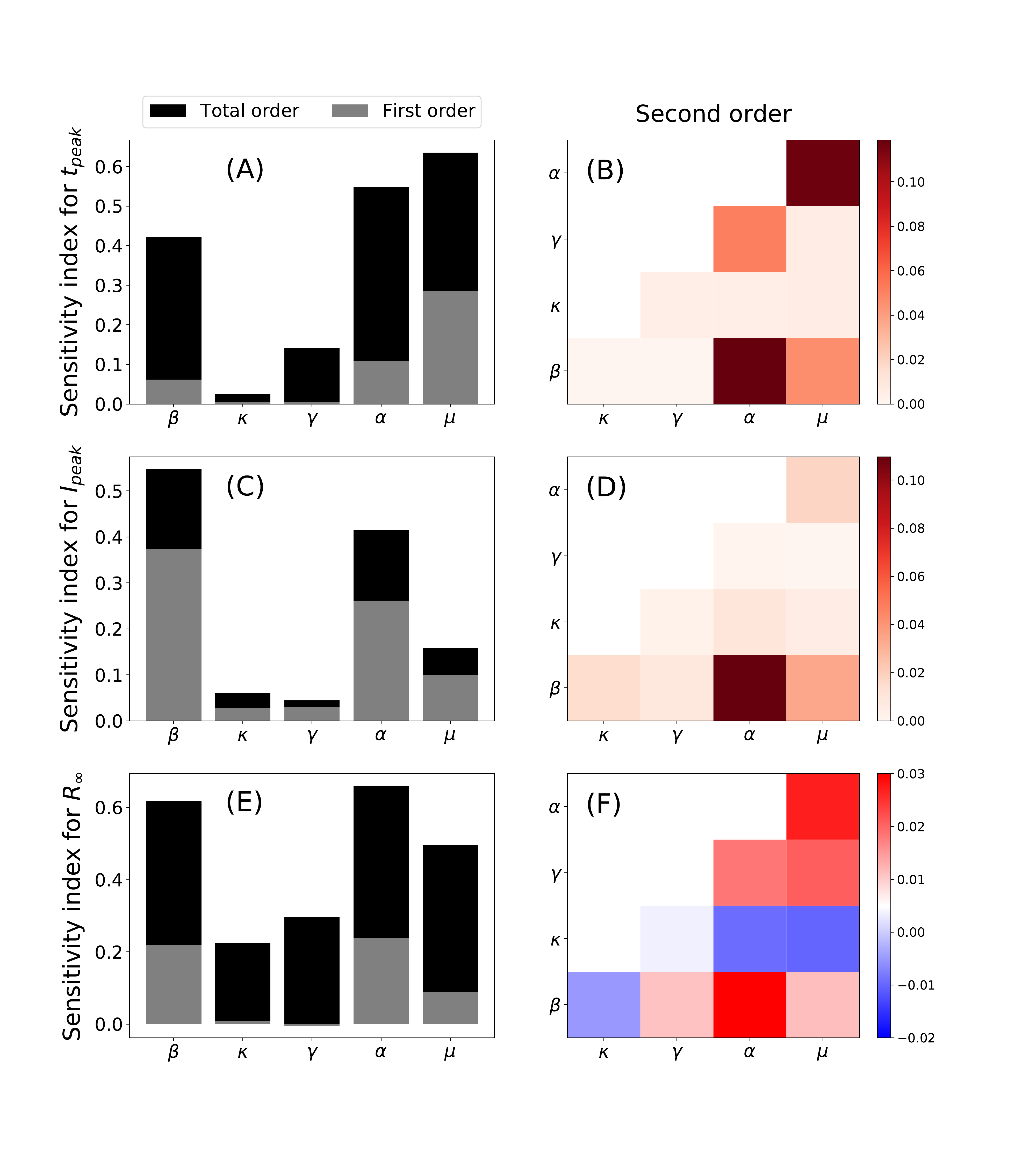}
        \caption{Global Sensitivity Analysis of the model parameters performed with the Sobol method with respect to the time at which the infected population peaks, $t_{peak}$ (A-B), the magnitude of this peak, $I_{peak}$ (C-D) and the final number of dead hosts, $R_{\infty}$ (E-F). The left column (A,C,E) shows the total and first-order indices and the right column (B,D,F) shows the second-order indices.}
        \label{fig:GSA}
    \end{figure}

    We computed the sensitivity indices for the model parameters with respect to the more relevant quantities of interest, namely, the time at which the number of infected hosts is maximal, $t_{\textrm{peak}}$, the maximum number of infected hosts, $I_{\textrm{peak}}$ and the final number of dead hosts, $R_\infty$. The results were obtained exploring the parameter space constrained to the intervals $\{\beta\in(0.001, 0.1), \ \tau_E\in(3,7), \ \tau_I\in(5,25), \ \alpha\in(0.001, 1), \ \mu\in(0.01, 0.04)\}$ using $10^4$ Quasi-Monte Carlo samples and are summarized in \cref{fig:GSA}.
    
    Parameters $\alpha, \ \beta$ and $\mu$ are the most influential with regard to the time at which the infected host population peaks, $t_{peak}$, the magnitude of the peak, $I_{peak}$, and the final number of dead hosts, $R_{\infty}$. The total output variance (total order indices) cannot be explained by the variances of the parameters alone (first order indices) (\cref{fig:GSA}). Therefore, higher-order interactions among the parameters importantly affect the sensitivity of the quantities under study. Indeed, the contribution to the total output variance of $\gamma$ and $\kappa$ for $t_{peak}$ and $R_{\infty}$ come notably from higher-order interactions. This can be checked in panels (B,C,F) of \cref{fig:GSA}, in which the contribution to the output variance from interactions between pairs of parameters (second order indices) is represented. Interactions among the parameters contribute to increasing the output variance with respect to $t_{peak}$ and $I_{peak}$, while the effect is more heterogeneous in the case of $R_{\infty}$. In particular, the interactions between $\alpha-\beta$ and $\alpha-\mu$ produce the main contributions to the increase of output variance in all cases, while $\kappa-\beta$, $\kappa-\alpha$ and $\kappa-\mu$ interactions decrease the output variance.

\subsection*{Epidemic control through vector management}

    The sensitivity analysis clearly indicates that acting on $\alpha,\beta$ and $\mu$ is the best strategy to lower disease incidence and mortality. However, controlling transmission rates is cumbersome so a different control strategy based only on vector control is considered in this section. In our model, there are two ways of implementing vector-population control: (i) decreasing the typical time, $1/\mu$, that vectors spend between crops each year by some mechanism (thus increasing $\mu$) and (ii) reducing the initial number of vectors that invade crops each year (e.g. lowering $N_v(0)$ via egg or nymph control (\cite{Lago2022}).

\begin{figure}[H]
        \centering
        \includegraphics[width=\textwidth]{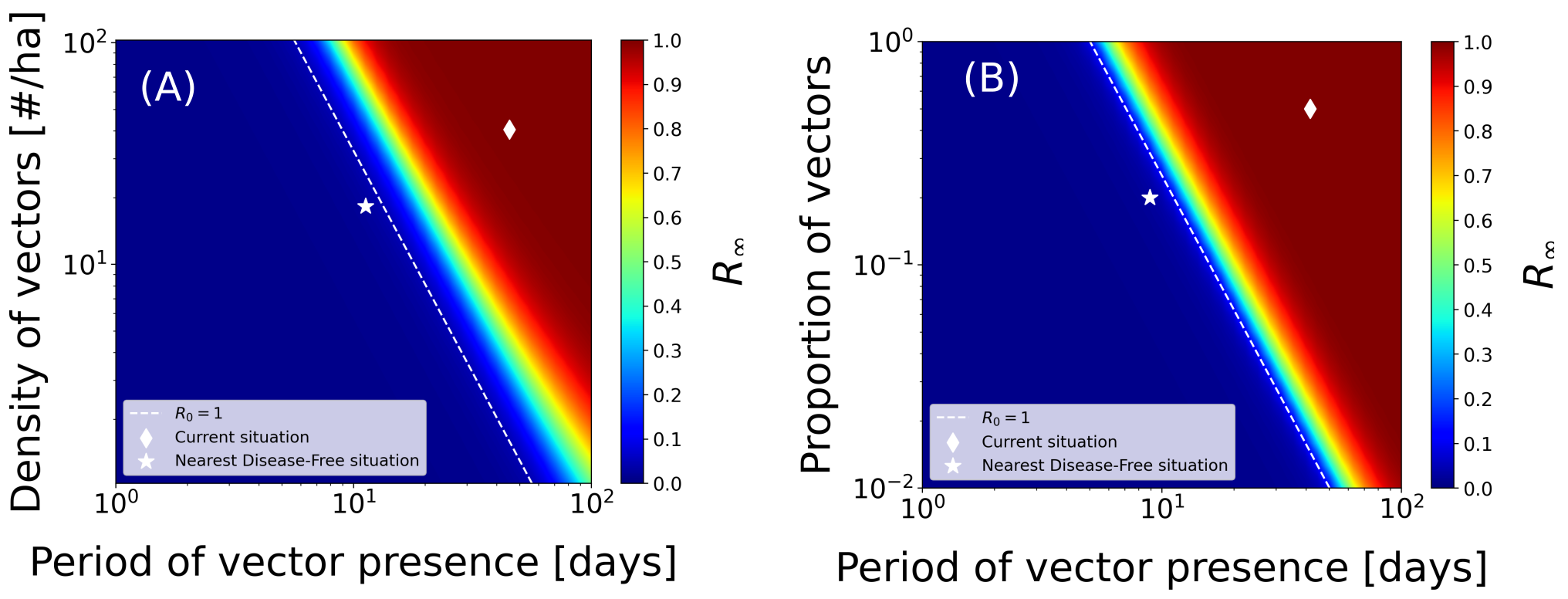}
        \caption{Epidemic control through vector management for ALSD in Mallorca (A) and OQDS in Apulia (B). The white shaded line denotes $R_0=1$, the white diamond corresponds to the parameter values of the fitted model. The white star is the closest disease-free state to the current situation in this representation.}
        \label{fig:control_strategy}
    \end{figure}

    We analyzed the effect of vector management by simulating epidemic outbreaks using different values of $\mu$ and $N_v(0)$, and keeping the rest of parameters as fitted for both ALSD and OQDS outbreaks (\cref{fig:control_strategy}). In both epidemics, decreasing the presence time as well as the number of vectors contribute to controlling the epidemic by lowering $R_0$ and, consequently, the final size of the epidemic, $R_{\infty}$. Furthermore, we observe that decreasing vector presence is more efficient than decreasing its annual initial population, i.e. we further reduce $R_{\infty}$, the final size of the epidemic, by applying a similar reduction in the residence time $1/\mu$. This could also be anticipated as $R_0$ depends quadratically on $1/\mu$ while only linearly on $N_v(0)$ (\cref{eq:R0}). However, the minimal intervention strategy, starting from the current situation in the $(1/\tau,N_v(0))$ parameter space that yields an absolute control of the epidemic, $R_0<1$, involves a mixed strategy of lowering both $1/\mu$ and $N_v(0)$.
    
\section*{DISCUSSION}

    In this work, we have developed a deterministic continuous-time compartmental model for \textit{Xylella fastidiosa} vector-borne diseases in Europe. The model attempts to characterize the main biotic processes that lead to the development of epidemics, including the seasonal dynamics of the main vector, \textit{P. spumarius}. We show how the model is sufficiently general to represent with some accuracy the parameters that determine the ALSD in Mallorca (Spain) and the OQDS in Apulia (Italy), both transmitted by \textit{P. spumarius}. To our best knowledge, this is the first mathematical model describing Xf epidemics that considers the temporal pattern of vector abundance observed in field data, faithfully representing the known biological information about the pathosystem. It includes a dynamic approximation of the non-stationary populations of \textit{P. spumarius}, mathematically represented by a sporadic source term through which vectors are born every year, and an exponential decay term. Due to the non-stationarity of the vector dynamics, $R_0$ in the model cannot be computed with standard methods such as the Next Generation Matrix (\cite{Diekmann2010}). To circumvent this problem, we applied an approximate method to compute it as previously proposed by (\cite{Gimenez2022}). We show that this approximate $R_0$ correctly characterizes the epidemic, further validating the method proposed by \cite{Gimenez2022}.
    
    Nonlinear mathematical models of disease transmission enhance our understanding of the different mechanisms operating in an epidemic, especially compared with correlative or machine learning methods, often very useful in practice but offering very little understanding. A key aspect to render these models useful is the determination of the parameters from available data. If this step can be properly performed, these models become very predictive and especially helpful to design disease control strategies. However, an appropriate calibration of the model relies on access to good-quality field data, which is often the bottleneck for the application of this kind of model. In the present study, the parameters have been obtained using a Bayesian inference framework, which relies on probability distributions rather than point-like measures. This way, mean or median values can be considered together with their confidence intervals able to characterize the robustness of the obtained parameters.
    
    One of the conclusions of the study is that the available data for both diseases is not enough to obtain robust estimates for all of the model parameters. The lack of data about the vector population compartments yields many possible values for the parameters that regulate transmission, $\alpha$ and $\beta$, provided that the progression of the host compartments correctly fits the field data. In other words, very infectious vectors (high $\beta$) that hardly ever get infected (low $\alpha$) can produce a similar outbreak within the host population to that produced by very low infectious vectors (low $\beta$) that get infected very often (high $\alpha$). The great difference in these situations would be that, in the former, the infected vector population would be very low, while in the latter, it would be quite high. This is a manifestation of parameter unidentifiability from the fit (\cite{Chowel2017,Roosa2019}), which stresses the importance of transmission and calls for detailed measurements of the vector population, and not just of the hosts. 
    
    In any case, our model shows that the vector-plant transmission process, mediated by $\beta$, is somehow different than the plant-vector one, mediated by $\alpha$. This heterogeneity can be caused by several factors: differences in the efficiency of plant-vector transmission with respect to vector-plant transmission, differences in contact rates, i.e. susceptible vectors contact trees at a different rate than infected vectors; vector feeding preferences, i.e. differences in the probability of contacting a susceptible host compared to an infected host, etc. Indeed, our mathematical model assumes constant contact rates with no preferences over any host state, so that under these assumptions, it indicates that the probability of effectively transmitting the pathogen from plant to vector is greater than from vector to plant. Furthermore, we found that the timing and magnitude of the infected host peak and the final number of dead hosts are mostly controlled by the vector-plant transmission rate, $\beta$, the plant-vector transmission rate, $\alpha$ and the vector removal rate $\mu$. Because these parameters are strongly related to the vector, the analysis makes clear that enhancing the knowledge about the vector, as well as obtaining precise data, is crucial to improve the modeling of Xf diseases and pose important questions to be solved in specifically designed experiments.
    
    The fact that the most influential parameters of the model are those related to the vector can be used to design appropriate disease control strategies. Because acting on transmission rates is rather cumbersome, we argue that control strategies should focus on reducing the vector population in crop fields. In our model, this depends on two parameters, $\mu$, the rate at which vectors exit the field and $N_v(0)$, the number of newborn susceptible vectors every year (assumed constant in this study). Our results show that a mixed strategy acting on both parameters is optimal to lower disease prevalence and, eventually, eradicate the disease. Interestingly, we also show that acting on the vector removal $\mu$ is more effective than controlling the newborn vector population $N_v(0)$. In fact, most control strategies carried out in practice for Xf diseases focus on the latter factor, reducing $N_v(0)$ via egg or nymph control (\cite{Cornara2018, lopez2022mechanical}). However, our results indicate that alternative strategies based on increasing the removal (or dispersal) rate of vectors should be explored. Furthermore, the evolution of the population compartments of the hosts and vectors provides relevant information on the epidemiology of both diseases. In both cases, the newly defined basic reproductive number that accounts for a decaying vector population is very predictive of the moment in which new infections are not produced anymore, coinciding approximately with the peak of infected hosts. Therefore, any intervention with control measures after this peak would have marginal effects on future disease progression.
    
    Our mathematical model is still rather simple, implementing only a few relevant epidemic processes in contrast to the high complexity of the pathogen-vector-host interactions occurring in plant epidemics. Indeed, the model itself raises some questions about these interactions, for example, whether or not contact rates are homogeneous. Another simplification of the model is the fact that the spatial constraints and the intrinsic stochasticity of the transmission processes are neglected. A straightforward extension of the model would be to include a specific spatial setting and implement the explicit motion of the vector within a stochastic framework, such as Individual Based Models (\cite{Grimm2005}). With this, the effectiveness of current and further control strategies could be tested and improved controlling for the motion of the vector. For instance, the control strategy based on the removal of symptomatic trees together with their surrounding trees at a given distance could be implemented in the model, evaluate the current effectiveness according to the present protocols and even provide improved parameters to be implemented in the field. Of course, implementing a model in which the spatial degrees of freedom are explicitly represented would require access to further information about vector mobility and spatially resolved data to confront the model, which is not currently available.
    
    Mathematical models tested against experimental data increase our understanding of the system under study. They also help to identify critical parameters that require better prior information to adjust functions relating to different variables and make the model predictions more accurate to suggest and test control strategies (\cite{cunniffe2015thirteen, jeger2018plant}). Our mathematical model suggests a certain lack of knowledge of the transmission processes and reveals that the currently available data is not enough to fit complex models dealing with the explicit dynamics of the vector population. \\
   
\noindent \textbf{Funding.} AGR and MAM acknowledge financial support from Grants No. RTI2018-095441-B-C22 (SuMaEco) and No. PID2021-123723OB-C22 (CYCLE) funded by MCIN/AEI/10.13039/501100011033 and by ``ERDF A way of making Europe'' and from Grant MDM-2017-0711 (María de Maeztu Program for Units of Excellence in R\&D) funded by MCIN/AEI/10.13039/501100011033.\\
 
\noindent The authors declare no conflict of interest.

\newpage
\printbibliography

\newpage
\appendix

\setcounter{figure}{0}
\makeatletter 
\renewcommand{\thefigure}{S\@arabic\c@figure}
\makeatother

\renewcommand{\thesection}{\Roman{section}}

\begin{center}
    \Large{\textbf{Appendices}}
\end{center}

\section{Vector population dynamics}\label{app:vector_dynamics}

\begin{figure}[H]
        \centering
        \includegraphics[width=1\textwidth]{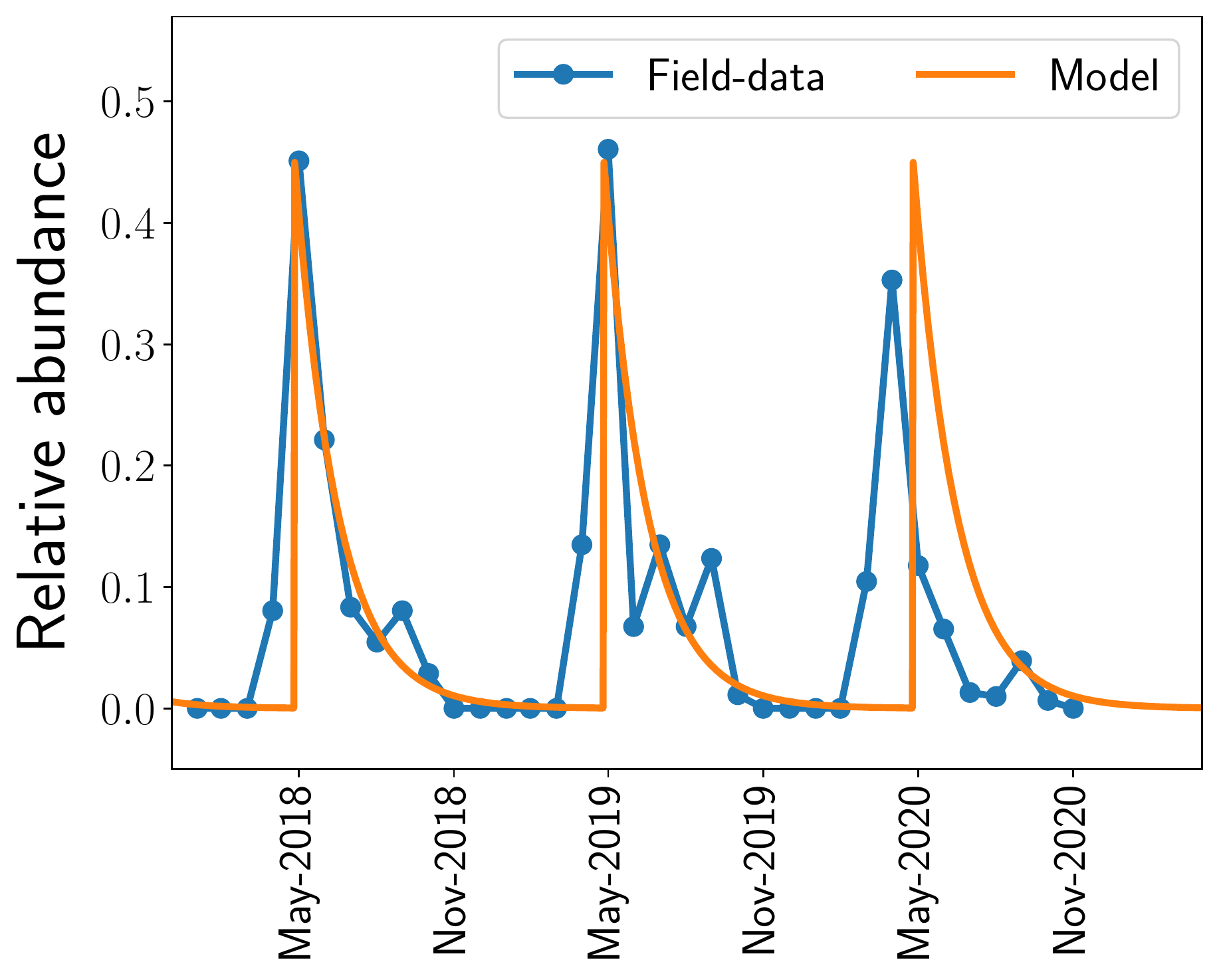}
        \caption{Vector dynamics produced by the model compared to field-data from (\cite{Lopez2021}).}
        \label{fig:vector_dynamics}
    \end{figure}

\cref{fig:vector_dynamics} shows a time series for the population of \textit{Philaenus spumarius} in Mallorca, taken from (\cite{Lopez2021}) (in blue). Superimposed (in orange) is the assumption used in our model \cref{eq:SEIR_v}, the $\delta(t-nT)$, i.e. every year susceptible vectors appear in the system.

\section{Determination of $R_0$}\label{app:R0}

    The handicap of determining the basic reproductive number of the model \cref{eq:SEIR_v} is that the pre-pandemic fixed point given by $I_H=I_v=0$ and $S_H=S_H(0)$ is not a fixed point of the system of differential equations, because vector population decays, so that the standard methods to compute $R_0$ such as the Next Generation Matrix (\cite{Diekmann2010, Gimenez2022}) do not apply. In (\cite{Gimenez2022}) a method was suggested to determine the basic reproductive number in the case of compartmental models of vector-borne transmitted diseases in which the vector population grows or decays. It consists in averaging the instantaneous basic reproductive number over the time of a generation.
    
    To proceed we consider that $I_H=I_v=0$, $S_H=S_H(0)$ is indeed a fixed point of the system. Then, the basic reproductive number could be determined, e.g. as shown in (\cite{Brauer2016}). First, an infectious host infects vectors at a rate $\beta S_H(0)/N_H$ for a time $1/\gamma$. This produces $\beta S_H(0)/\gamma N_H$ infected vectors. The second stage is that these infectious vectors infect hosts at a rate $\alpha N_v(0)/N_H$ for a time $1/\mu$, producing $\alpha N_v/\mu N_H$ infected hosts per vector. The net result of these two stages is
    \begin{equation}
        \tilde{R}_0=\frac{\alpha\beta}{\mu\gamma}\frac{S_H(0)}{N_H^2}N_v(0)=R_0^* \cdot N_v(0)\ .
        \label{eq:R0tilde}
    \end{equation}
    This result coincides with the value of $R_0$ obtained using the standard NGM method, that can be applied in this case because we are assuming that we use a nongeneric initial condition that sits at the fixed point of the model. 

    In practice, our initial condition will never be a fixed point of the model, and, as mentioned above, we will obtain an approximate basic reproductive number, to which we will refer as $R_0$ using the method suggested in (\cite{Gimenez2022}), that consists in (\cite{Gimenez2022}). calculating the \textit{average} number of secondary infections produced by an infected host in \textit{one generation}. First defines an instantaneous basic reproductive number,
    \begin{equation}\label{eq:R0i}
        R_0^{(i)}(t)=\frac{\beta\alpha}{\mu\gamma}\frac{S_H(0)}{{N_H}^2} N_v(t)=R_0^* N_v(t) \ ,
    \end{equation}
    from which the average is simply computed as
    \begin{equation}\label{eq:R_eff_to_integrate}
        R_0=\avg{R_0^{(i)}(t)}\Big\rvert\limitss{0}{\tau}=R_0^*\avg{N_v(t)}\Big\rvert\limitss{0}{\tau}=R_0^*\frac{1}{\tau}\int_0^{\tau}N_v(t) \, \dif t \ ,
    \end{equation}
   In our model, the time-dependent vector population can be obtained from \cref{eq:SEIR_v},
   \begin{equation}
       \dot{N}_v=\dot{S}_v+\dot{I}_v=-\mu N_v \Longrightarrow N_v(t)=N_v(0)e^{-\mu t} \ ,
   \end{equation}
    and introducing this expression for $N_v(t)$ in \cref{eq:R_eff_to_integrate} the integral can be solved
    \begin{equation}
        R_0={{\beta\alpha S_H(0)}\over{\mu\gamma N_H ^2}}
        \,\frac{N_v(0)}{\mu\tau}\left(1-e^{-\mu\tau}\right)=
        R_0^*\ \frac{N_v(0)}{\mu\tau}\left(1-e^{-\mu\tau}\right) ,
        \label{eq:r0mdef}
    \end{equation}
    that is an approximated expression to the basic reproductive number for our model, in which the vector population is nonstationary,
    where, in \cref{eq:R0i} and \cref{eq:r0mdef} it has been defined, $R_0^*= (\beta\alpha S_H(0))/(\mu\gamma N_H ^2)$.
    
    Note that in our model one generation correspond to one year and that $N_v(0)$ is reset every year.





\end{document}